\documentclass[useAMS,usenatbib]{mn2e}
\usepackage{graphics,epsfig}

\def \be{\begin{equation}}
\def \ee{\end{equation}}
\def \bea{\begin{eqnarray}}
\def \eea{\end{eqnarray}}
\def\etal{{et al.\ }}

\title[Extended X-ray emission from radio galaxy cocoons]{
Extended X-ray emission from radio galaxy cocoons 
}

\author[Biman B. Nath]
{Biman B. Nath\thanks{biman@rri.res.in}\\
Raman Research Institute, Sadashiva Nagar, Bangalore 560080, India\\
}

\begin{document}
\date{Accepted 2010 May 19.
      Received 2010 May 19;
      in original form 2010 February 13}

\maketitle

\label{firstpage}

\begin{abstract}
We study the emission of X-rays from lobes of FR-II radio galaxies by
inverse Compton scattering of microwave background photons. Using a
simple model that takes into account injection of relativistic electrons, 
their energy losses through adiabatic expansion, synchrotron and inverse
Compton emission, and also the stopping of the jet after a certain
time, we study the evolution of the total X-ray power, the
surface brightness, angular size of the X-ray bright region and the X-ray
photon index, as functions of time and cocoon size, 
and compare the predictions with observations.
We find that the radio power drops rapidly after
the stopping of the jet, with a shorter time-scale than the X-ray power.
The X-ray spectrum initially hardens until the jet stops because the
steepening of electron spectrum is mitigated by the
 injection of fresh particles,
for electrons with $\gamma \ge 10^3$. This happens because of the 
concurrence of two times scales, that of the typical jet lifetimes
and cooling due to inverse Compton scattering ($\sim 10^{7\hbox{--}8}$ yr),
of electrons responsible for scattering CMB photons
into keV range photons (with $\gamma \sim \sqrt{1 \, {\rm keV}/ kT_{CMB}}$).
Another finding is that
the ratio of the X-ray to radio power is a robust parameter that varies 
mostly with redshift and ambient density, but is weakly dependent on other
parameters.
We also determine the time-averaged ratio of X-ray to radio luminosities
(at 1 keV and 151 MHz) and find that it scales with redshift as $\propto
(1+z)^{3.8}$, for typical values of parameters. 
We then estimate the X-ray luminosity function of FR-II
radio galaxies and estimate the number of these diffuse X-ray bright objects
above a flux limit of $\sim 3 \times 10^{-16}$ erg cm$^{-2}$ s$^{-1}$ to be
$\sim 25$ deg$^{-2}$.
\end{abstract}

\begin{keywords}
galaxies: active -- galaxies: intergalactic medium -- X-rays: galaxies
\end{keywords}

\section{Introduction}
A number of radio galaxies have been observed to emit diffuse X-rays
in the region between the nucleus and radio hot spots in recent
years. This emission has been interpreted as inverse-Compton
scattering of the cosmic microwave background (CMB) photons by
relativistic electrons in the radio lobes that have lost most of their
energy and their Lorentz factor have come down to $\gamma \sim 10^3$
(Fabian \etal 2003; Blundell \etal 2006; Erlund \etal 2008; Johnson
\etal 2007; Croston \etal 2005; Fabian \etal 2009; Isobe \etal 2009). 
That relativistic electrons in radio
galaxy cocoons could upscatter CMB photons to X-rays has long been anticipated
since the discovery of CMB (e.g., Felten \& Rees 1967).

Since the increase in the CMB energy density
with redshift ($\propto (1+z)^4$) compensates for the cosmological
surface brightness dimming, this extended emission has been seen at both
low and high redshift, and has been used as a probe of the relativistic
plasma in the lobe.  Combined with radio observations, this emission
provides constraints on the magnetic field and on the energy distribution
among the relativistic particles, or both. 
While electrons with $\gamma \ge10^4$ to emit GHz synchrotron radiation
(for typical magnetic fields of a few $\mu$G), it requires $\gamma \sim 10^3$
to upscatter the CMB photons to the observed $1\hbox{--}10$ keV range in
X-rays.

The simultaneous radiation in two different wavelengths caused by the
same parent electron population provides an excellent opportunity to 
probe the physical nature of the radiating object.
Comparing the X-ray and radio emission from lobes,
Blundell \etal (2006) showed that the injected electron energy
distribution have a low-energy turn-over at $\gamma _{min} \ge 10^3$,
since the X-ray bright regions did not coincide with the radio hot spots.
The comparison between X-ray and radio studies also shed light on the magnetic
field.
Croston \etal (2005) argued
that the inferred magnetic field strength is close to the equipartition
value (lying between $0.3\hbox{--}1.3$ times the equipartition value). 

In this paper, we use a variant of
a model of emission from FRII radio galaxies advocated
by Kaiser \& Alexander (1997) (hereafter KA97)
 and Kaiser, Dennett-Thorpe \& Alexander (1997)
(hereafter KDA97) to study the evolution of X-ray power and surface
brightness with time, and as functions of the radio lobe size, ambient
density, redshift and the jet power. We do not assume self-similarity
of the evolution of the radio lobe assumed by KA97 and KDA97, in order
to study the effect of the cessation of jet activity on the X-ray
surface brightness.

The evolution of radio lobes after the stopping of jets has been 
discussed
by Komissarov \& Gubanov (1994), although they did not address the soft
X-ray emission by old electrons. Kaiser \& Cotter (2002) have also 
discussed the evolution of radio galaxies in connection with the observations
of radio relics, and Reynolds \etal ( 2002) have numerically
 studied the expansion of
'dead' radio galaxies.

We first discuss the KA97 and KDA97 model and the changes introduced in
the model in the present paper, and then discuss the results of our 
calculations.

\section{Evolution of radio galaxy lobes for FRII sources}
\subsection{The KDA97 model}
The KDA97 model of the dynamical expansion of an FR II-type radio source
assumes a self-similar expansion, driven by twin jets emerging from the
nucleus in opposite directions, pushing the surrounding environment. The
jets produce strong shocks where the jet particles are accelerated and
these particles made the cocoon expand. The density distribution in the
ambient medium is assumed to be a power-law, $\rho(r)=\rho_0 (r/a_0)^{-\beta}$,
where $\rho_0$ is the density at a core radius $a_0$. If the jet power
is denoted by $Q_j$, then the combination $(Q_j/\rho_0 a_0^\beta)^
{{1 \over 5 -\beta}} \, t^{{3 \over 5 - \beta}}$ 
has the dimension of length, and, in a self-similar model,
it is proportional to the
length of the jet, $L_j (t)$ (Falle 1991). Half of the cocoon volume is 
approximated by
a cylinder of length $L_j$ and a base radius.
 The ratio $R$ between the length and 
half-width of this cylinder is referred to as its axial ratio.

The dynamics of the cocoon is determined by the cocoon pressure, $p_c
=(\Gamma_c-1)(u_e+u_B)$, where the contributions from different components
have been added: (a) relativistic electrons with energy density $u_e$ and
adiabatic index $\Gamma_c$ and (b) a tangled magnetic field with energy
density $u_B$. (This implicitly assumes that $\Gamma_c=4/3$, the 
adiabatic index appropriate for magnetic field pressure.) Suppose the 
cocoon volume scales as $t^{a_1}$.
Then the cocoon pressure evolves with time
as $p_c \propto t^{- \Gamma_c a_1}$ (see KDA97). The cocoon is assumed to 
be divided into small volume segments filled with magnetized plasma
and particles that are injected into the cocoon at the jet terminal
shock.

The initial
electron
energy distribution is assumed to be a power law in energy, 
\be
n(\gamma_i,t_i) d\gamma=n_0 \gamma_i^{-p} d\gamma_i
\ee
where the spectrum ranges between $\gamma_{min}$ and $\gamma_{max}$.
The electrons lose energy through adiabatic expansion, synchrotron
and inverse Compton losses.
KDA97 found a closed-form solution for
 the energy distribution
at a later time $n(\gamma, t)$ in the case of 
a self-similar expansion of the cocoon.

The ratio of the energy density in particles to that in magnetic
field is assumed to be $r$, so that the 
magnetic energy density $u_B$  at time $t$ is given by,
\be
u_B(t)={r p_c (t) \over (\Gamma_c -1 ) (r+1) } \,.
\label{eq:magfrac}
\ee
Assuming that for synchrotron radiation,
electrons emit only at their critical frequency $\nu=\gamma^2 \gamma_L$,
where $\nu_L$ is the Larmor frequency, the total radio
power emitted by electrons in a volume element $\delta V$ of the cocoon
(radio power per unit frequency and solid angle) is given by,
\be
\delta P_\nu= {1 \over 6 \pi} \sigma_T \, c \, 
u_B {\gamma ^3 \over \nu} n(\gamma) 
\delta V \,. \nonumber\\
\label{eq:syn}
\ee
Here the volume element at time $t$ is related to the initial time
interval over which the particles residing in it were injected, as,
\be
\delta V(t)={(\Gamma_c -1) Q_j \over p_c(t_i)} (4 R^2)^{(1-\Gamma_c)/\Gamma_c}
\, \Bigl ( { t \over t_i } \Bigr)^{a_1} \, \delta t_i \,.
\label{eq:kdavol}
\ee
The total radio power is then calculated by integrating over all
time,
\be
P_\nu=\int_0 ^t \delta P_\nu \,,
\label{eq:synint}
\ee
using eqns (\ref{eq:syn}) and (\ref{eq:kdavol}).

\subsection{Evolution without self-similarity}
In this paper, we would like to study the evolution of radio and X-ray
power on a time scale that is longer than the typical jet lifetime, and study
the effect of the stopping of jets on the X-ray emission. 
As we will explain later in the paper, the stopping of the jet has important
effects on X-ray power of the cocoon, and the results are markedly 
different than in the case where the jet continues to be active. To do this,
we cannot use the self-similar evolution assumed in KDA97. Instead, we
calculate the size evolution the radio lobe in the following way that 
captures the basic physical processes in a simple manner. 

Following the standard evolutionary picture (Scheuer 1974; Begelman \& Cioffi
1989, Reynolds \& Begelman 1997), that radio jets inject
relativistic particles into a cocoon, which is overpressured
(Nath 1995) and drive a strong shock into the ambient medium. The speed 
with which this shocked shell expands is determined by the ram pressure
of the ambient gas that enters the shock. Assuming that the pressure inside
the cocoon at any given time is uniform (see KA97), and
a value $p_c(t)$, and assuming that radiation loss is small, we have
for the expansion of the cocoon,
\bea
Q_j (t)&=& {1 \over \Gamma_c -1} (V_c \dot{p_c} + \Gamma_c p_c \dot{V_c})
\nonumber\\
{dL_j \over dt}&=&\sqrt{{p_c \over \rho_0}} \,,
\label{eq:dyn}
\eea
where $V_c$ is the volume of the cocoon, and $\rho_0$ is the ambient density.
For simplicity, we assume a constant axial ratio $R$,
and the volume is assumed to be, $V_c={ \pi \over 4 R^2} L_j^3$ (KA97).

For the jet luminosity we specify a jet lifetime $t_j$, so that
$Q_j$ is a constant for $t\le t_j$, and $Q_j=0$ for $t> t_j$, where
$t_j \sim 10^{7\hbox{--}8}$ yr. Numerical simulations of 'dead'
radio galaxies after the cessation of jet activity have shown that 
the overpressured cocoon continues to expand until a pressure equilibrium
is established with the ambient medium (Reynolds, Heinz, Begelman 2002).

We assume a  constant axial ratio $R\sim 2$, which is an average value (Leahy
and Williams 1984), and, following Wang \& Kaiser (2008), 
we use a profile of the ambient density as given 
by,
\be
\rho_a=\rho_0 (r/a_0)^{-\beta}=\Lambda r^{-\beta} \,,
\ee
where $\Lambda=\rho_0 a_0^\beta$. For $\beta=2$, $\Lambda$ has the units
of [g cm$^{-1}$], and one can infer its value from the observations of 
environments of radio galaxies. Typically, for $\beta=2$ and $a_0\sim
200$ kpc, an electron density $n_{e,0}(\equiv \rho_0/(\mu m_p)
 \sim 10^{-4}$ cm$^{-3}$ would
imply $\Lambda \sim 10^{19}$ g cm$^{-1}$, a value we assume for most of
our models in this paper. 
Observations of individual galaxies  show $\Lambda$
to be in the range of $10^{19\hbox{--}20}$
g cm$^{-1}$ (Fukazawa \etal 2004), whereas Jetha \etal (2007) 
inferred $\Lambda \sim 10^{19}\hbox{--} 10^{20}$ g cm$^{-1}$
in group environments, although these determinations assume $\beta \sim 1.5$. 
For simplicity, because of the ability of combining two free parameters
($\rho_0$ and $a_0$) into one ($\Lambda$), we assume a value of $\beta=2$
for our calculations.

Solving equations \ref{eq:dyn}, we can therefore determine the cocoon pressure
$p_c(t)$ at any instant, from which we determine an equipartition
value of magnetic field, using eqn (\ref{eq:magfrac}).
assuming (as in  KDA97) a ratio $r$ between
particle and magnetic field energy density.
The corresponding electron energy density is then given by
$u_e(t)= u_B(t)/r$, assuming
that there is no thermal particles in the cocoon
(i.e., in the language of KDA97, $u_T=0$). 

In the KDA97 model, particles are assumed to be injected at the termination
point of the jet. In our case, after the jet stops, the injection of particles 
also stops, and so the integration in the equation (\ref{eq:synint}),
for the determination of radio power at a given time $t$, 
has an upper limit of $\min(t,t_j)$.
To perform this integration,
we rewrite the equation
for volume segments (eqn (\ref{eq:kdavol})) which in the
KDA97 model used the exponent $a_1$ for self-similar evolution of cocoons.
For $t \le t_j$,
any small volume segment of the cocoon at a a given instant $t$ can be
related to the pressure at the time ($t_i$) when the electrons in this segment
were injected into the cocoon. One can 
rewrite equation (\ref{eq:kdavol}) in the KDA97 model as (for $t<t_j$)
\be
\delta V(t)={(\Gamma_c -1) Q_j \over p_c(t_i)} (4 R^2)^{(1-\Gamma_c)/\Gamma_c}
\, \Bigl ( { p_c(t) \over p_c(t_i) } \Bigr)^{1/\Gamma_c} \, \delta t_i \,.
\label{eq:kdavol2}
\ee

The evolution of Lorentz factor is explicitly solved using the loss
equation,
\be
{d \gamma \over dt}=-{1 \over 3} {1 \over V_c} {dV_c \over dt} -{4 \over 3} 
{\sigma_T \over m_e c} \gamma ^2 (u_B+u_c) \,,
\label{eq:gamma}
\ee
where the first term denotes adiabatic energy loss and is computed
using the results of equations (\ref{eq:dyn})). The second term
combines radiation loss in synchrotron and inverse-Compton scattering.
Here $u_c=aT_{CMB}^4$ is the CMB photon energy density, $
\sigma_T$ is the Thomson cross-section, $m_e$ is the electron mass
and $c$ is the speed of light. 

Thus, beginning with an initial energy distribution law with a power-law
index $p$, one can solve for the energy distribution at any given time,
$n(\gamma, t)$, given the initial distribution, $n(\gamma_i, t_i)$. Note
that this was analytically done by KDA97 for a self-similar evolution
of the cocoon, and we explicitly solve it in order to go
beyond self-similarity.

Using this knowledge of $n(\gamma, t)$, we can then 
use equation (\ref{eq:syn}) in conjunction with eqn (\ref{eq:kdavol2}),
and integrate over time to calculate the radio power 
at a given frequency and at a given instant.

\subsection{Inverse-Compton radiation}
We extend this formalism further to determine the inverse-Compton 
(IC) emission as a function of time. Firstly, we note that electrons with
Lorentz factor $\gamma$ boosts a CMB photon of frequency $\nu_{CMB}$
into an energy $h \gamma^2 \nu_{CMB}$. The precise calculation of the
IC power for this photon would require one to consider the total spectrum
of CMB photons. But we can simplify it for our purpose here by assuming
all CMB photons to have a single frequency. We assume that the CMB photon
distribution function is given by $v'(\epsilon)=v'_0 \delta (\epsilon -
k_BT_{CMB})$, where the normalizing factor $v'_0$ can be calculated by
requiring that total energy density $\int \epsilon v'(\epsilon) \, d\epsilon
=aT_{CMB}^4$, and one has $v'_0=aT_{CMB}^3/k_B$. 
The total scattered power depends on the integral (see equation (7.29a) in
Rybicki \& Lightman 1979),
\be
I'=\int d\epsilon \, v'(\epsilon) \, \epsilon^{{ p-1 \over 2}} 
={a T_{CMB} ^3 \over
k_B} (k_B T_{CMB})^{(p-1)/2} \,.
\ee
If one had used the blackbody
 distribution function $v(\epsilon)=(8 \pi \epsilon^2 /
h^3 c^3) (\exp( \epsilon /k_BT_{CMB}) -1)^{-1}$, then the corresponding
integral would have yielded (see equation 7.31 in Rybicki and Lightman 1979),
\be
I=\int d\epsilon \, v(\epsilon) \, \epsilon^{{p-1 \over 2}} =
{8 \pi \over h^3 c^3} (k_B T_{CMB})^{{p+5 \over 2}} \, \Gamma ({p+5 \over 2})
\, \zeta ({p + 5 \over 2}) \,,
\ee 
where the symbols have the standard meanings. Then the ratio,
\be
{I' \over I} ={ \pi ^4 \over 15} {1 \over \Gamma ({p+5 \over 2})
\, \zeta ({p + 5 \over 2})} \,,
\ee
shows the error one incurs in assuming all CMB photons to have the peak
frequency in estimating the inverse Compton power.
For $p=2.2$, one finds
$I'/I \sim 1.6$. Therefore the total scattered power 
calculated using this assumption is correct within an accuracy of 60\%, and we
adopt it for simplicity in our calculation.

We therefore calculate the IC power at a given frequency, in the manner
of previous eqn (\ref{eq:syn}), as,
\be
P_{IC, \nu}=\int_0 ^{\min [t,t_j]} 
{1 \over 6 \pi} \sigma_T c u_c {\gamma ^3 \over \nu} n(\gamma) 
\delta V \,,
\ee
using eqn (\ref{eq:kdavol2}) for $\delta V$, and the value of
$n(\gamma, t)$) found from solving eqn (\ref{eq:gamma}).

Not all volume elements in the cocoon would contribute to the
IC radiated power though. Electrons in some volume element 
that was injected at an earlier time $t_i$  may lose energy to the extent
that they fail to boost CMB photons to the observed X-ray band. 
These volume elements would not contribute to the X-ray power. 

We calculate the total projected
area of the cocoons that contribute to X-ray power,
assuming the cocoon to be in the plane of the sky, and 
an axial ratio $R$.
This allows us to calculate the X-ray surface brightness.
We also calculate the fraction $f_x$ of the total surface area of the
cocoon that contributes  to X-ray emission.

\section{Results}
To begin with, as a fiducial case, we consider $Q_j=10^{45}$ erg s$^{-1}$,
$\Lambda=10^{19}$ g cm$^{-1}$ 
at $z=0.1$, and a constant axial
ratio $R=2$. Following KDA97, we also adopt
$r=0.785$ and $p=2.14$, $\gamma_{max}=10^6$ and $\gamma_{min}=1$.
We refer to this set of parameters as Case I.

\begin{figure}
\centerline{
\epsfxsize=0.5\textwidth
\epsfbox{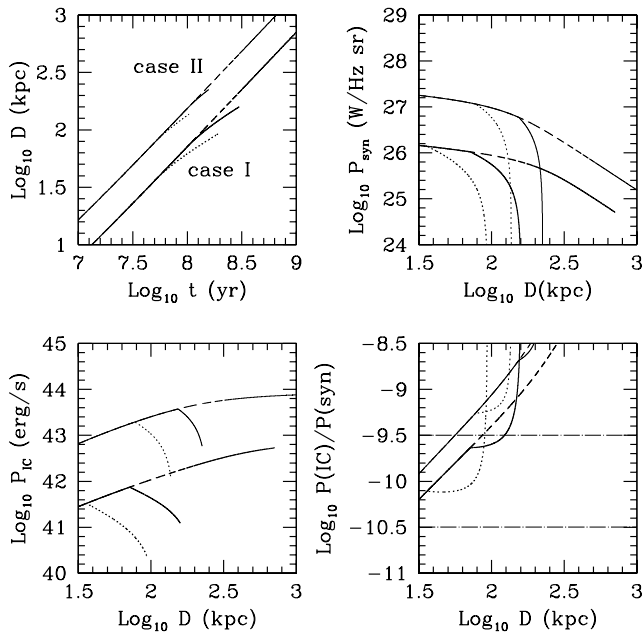}
}
{\vskip-3mm}
\caption{
Time evolution of radio and X-ray powers are shown in this plot.
Thick lines show the result for Case I ($Q_j=10^{45}$ erg
s$^{-1}$, $\Lambda=10^{19}$ g cm$^{-1}$ at $z=0.1$, with $\gamma_{min}=1$),
whereas thin lines show the results for Case II ( $Q_j=10^{46}$ erg
s$^{-1}$ at $z=1$, with the same ambient density as in Case I).
Dotted, solid and
dashed lines show the cases for $t_{j}=5 \times 10^7, \,
10^8 $ and $10^9$ yr, respectively. 
Time evolution of the cocoon size ($L_j$) is shown in the top-left panel,
radio power at 178 MHz (in W/Hz sr) in the top-right, and the
inverse Compton power at 1 keV in the same units is shown
in the bottom -left panel.
The bottom right panel shows
the corresponding ratio of inverse Compton to radio power as a function
of cocoon size for the three cases. 
Two horizontal dot-dashed lines indicate the range of observed ratios
(see text).
}
\end{figure}

The set of thick lines in Figure 1 show the results
for this fiducial case. We
plot the time evolution of cocoon size in the top-left panel,
radio power at 178 MHz in the top-right panel, IC power at 1 keV in the
bottom-left panel.
The corresponding evolution of the ratio of synchrotron to IC power
with cocoon size is shown in the bottom-right panel.
The solid lines show  the case for $t_j=10^8$ yr, and dotted and dashed lines
show the cases $t_{j}=5 \times 10^7$
and $10^9$ yr, respectively.

In the same figure, we also show the results of another case, with
a higher jet power and one located at a higher redshift than in the
fiducial case. The thin set of lines in Figure 1 show the case with
$Q_j=10^{46}$ erg s$^{-1}$ at $z=1$ (Case II), 
keeping the values of other parameters
the same.

The evolution of the size of the cocoon is self-similar, as expected, till
$t \sim t_j$ (top left panel of Figure 1), after which the cocoon size
expands slower than before. The radio luminosity as a function of time
(top-right panel)
is comparable to the results of KDA97 for the case of $t_j \sim 10^9$ yr, with
sources at higher redshift dimming faster than their low redshift counterparts
because of inverse Compton loss. For
shorter jet lifetime, the radio luminosity drops precipitously after $t
\sim t_j$. This is mainly driven by rapidly declining pressure inside
the cocoon owing to adiabatic expansion.

The X-ray luminosity of the cocoon at 1 keV (bottom-left panel of Figure 1)
increases till $t\sim t_j$, as high energy electrons lose energy ($dE/dt \propto
E^2$) and enter the energy bins that are conducive for inverse Compton 
scattering CMB photons to the soft X-ray band. 
The number of electrons with Lorentz factors $\sim 10^3$ increase
because the draining of these electrons to lower energy bins is more
than compensated by the supply of energetic electrons from the jet.
But after the jet stops, no new electrons are supplied to the cocoon, and
the X-ray power drops rapidly (more so at high redshifts). 

The time scale
for inverse Compton loss is  $t_{ic}
 \sim 3 \times 10^7$ yr for $\gamma \sim 10^3$
at present epoch. The time scale for the drop in X-ray power is made
shorter than this by the additional loss of energy from adiabatic expansion
of the cocoon. It is clear from curves in Figure 1
 that the stopping of the jet activity has a 
profound effect on the X-ray power that would have been missed in a 
calculation with self-similar evolution of cocoons. Self-similar models
would have predicted a rising X-ray power, although the rate of increase
would have tapered beyond a time-scale $\sim t_{ic}$, because the loss
of electron energy would have been continually compensated by injection
of new and energetic electrons. In the case of jet stopping after $t_j$,
this privilege is withdrawn from the cocoons and their X-ray power
rapidly diminishes with time.  

The ratio of X-ray to radio power (bottom-right panel), 
therefore, shows a mild increase till $t_j$,
after which it increases rapidly, because of the drop in radio luminosities.
We note that, typically the value of the ratio at $t \sim t_j$ is of order
$P_{ic}/P_{s} \sim 10^{-10}\hbox{--}10^{-9}$. For a single power law
electron energy
distribution with index $p=2\alpha +1$, where $\alpha$ is the synchrotron
(and inverse Compton) radiation spectral index, the ratio of 
relative luminosities can be written in terms of the strength of the magnetic
field and the CMB temperature. For $p=2.5$, we have (e.g. Tucker 1977),
\bea
{P_{ic} \over P_s}\approx
4.7 \times 10^{-11} \Bigl ({T_{CMB} \over 2.73 \, K} \Bigr )^{3 + \alpha} \,
\Bigl ({B \over 10 \, \mu G} \Bigr )^{-(\alpha +1)} \,
\eea
where the coefficient is $\sim 2.7 \times 10^{-11}$ for $p=2$, and 
$8.3 \times 10^{-12}$ for $p=3$. These values are somewhat lower than
predicted from our detailed calculations, which take into account the
electron energy distribution in different parts of the cocoon as they
evolve. Although the magnetic field strength throughout the cocoon
is assumed to the same at any given time, the difference in the evolution
of electron energy spectrum in different parts of the cocoon can give
rise to a value of the ratio $P_x/P_r$ that is different from the above simple
estimate. Admittedly, the flux ratio is mainly driven by magnetic field,
which is what the above simple estimate indicates, but our calculations 
show  the important role played by the evolution of electron spectrum 
 prior to the observed epoch for an object.


The observed ratios of fluxes at 1 kev and 178 MHz fall in the range of
$P_x/P_r \sim 10^{-10.5} \hbox{--} 10^{-9.5}$, for objects with a variety
of linear sizes and at different redshifts (e.g., 
3C 215 ($z=0.41$), 3C 334 ($z=0.56$), 3C 9 ($z=2.0$) (Bridle \etal 1994),
3C 219 ($z=0.17$) (Perley \etal 1980), 3C 265 ($z=0.81$) (Bondi \etal 2004),
3C 109 ($z=0.31$) (Giovannini \etal 1994), 3C 179 ($z=0.85$) (Sambruna
\etal 2002)). We indicate this range by two dot-dashed lines in the figure.
Since the evolution of this ratio  depends on many parameters (as listed in 
Table 1, for example), our goal in this paper is not to reproduce all the 
observed parameters, but to probe whether or not the model yields
 representative values consistent with observations.


It is apparent from the figure that the observed range of ratios are mostly
consistent with model predictions for different parameters. It also appears
that the
observed range is {\it not} related to the rapidly rising part of the 
evolutionary curves. In other words, the systems observed
in X-ray (near 1 keV) should also be radio bright at low frequencies,
of order $\sim 150$ MHz, that we have used for our calculations ($\sim
 178$ MHz). Low frequency
radio observations therefore should be an important probe of X-ray
bright cocoons, as has been pointed out by previous authors.

Another aspect is that cocoons spend a relatively short time in the rapidly
rising parts of the curves for ratio between $P_x/P_r$, because of the
precipitous drop in radio luminosity after $t_j$. In other words, 
radio galaxies that have faded from the window of frequencies of
order a few hundred MHz, are unlikely to be X-ray bright, since the 
X-ray power would also drop (see bottom-left panel of Figure 1).

\begin{figure}
\centerline{
\epsfxsize=0.5\textwidth
\epsfbox{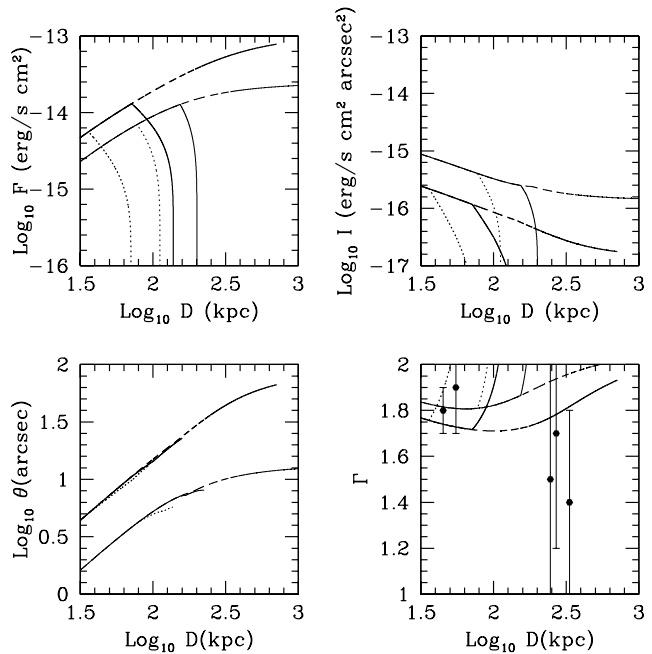}
}
{\vskip-3mm}
\caption{
Evolutions of X-ray flux in the 1-5 keV band (top-left panel), X-ray
surface brightness (top-right), angular size of the X-ray bright region
(bottom-left) are shown as functions of the cocoon length, for the same
cases as in Figure 1. Thick lines refer to the fiducial case of
 $Q_j=10^{45}$ erg
s$^{-1}$, $\Lambda=10^{19}$ g cm$^{-1}$ at $z=0.1$, with $\gamma_{min}=1$
(Case I).
Thin lines consider $Q_j=10^{46}$ erg
s$^{-1}$ at $z=1$ (Case II).
 Solid, dotted and
dashed lines show the cases for $t_{j}=5 \times 10^7, \,
\times 10^8 $ and $10^9$ yr, respectively. The bottom right panel shows
the corresponding X-ray photon index in the 1-5 keV band as a function
of cocoon size for these cases. We also plot the data points for
3C 47N ($z=0.43$, $L_j\sim 333$ kpc),
3C 215 ($z=0.41$, $L_j\sim 246$ kpc) (Bridle \etal 1994),
3C 219 ($z=0.17$, $L_j\sim 270$ kpc), 3C 452 ($z=0.81$, 
$L_j\sim 45$ kpc) (Perley \etal 1980), 
3C 265E ($z=0.81$, $L_j\sim 55$ kpc) (Bondi \etal 2004),
for comparison.
}
\end{figure}

Next, we plot  in Figure 2 the corresponding flux in the 1-5 keV band
(top left panel) for the cases shown in Figure 1. 
Our results show that for 
the parameters used here the cocoons typically
emit at a flux of $\sim 10^{-14} $ erg s${-1}$ cm$^{-2}$
per keV (or a few micro-Crab)
before dropping after $t_j$, consistent with observed fluxes (see, e.g.,
Laskar \etal 2010).

We also need to consider the X-ray surface brightness apart from the total flux.
We calculate the total
area of the X-ray bright region of the cocoon, by summing over the volume
elements in which electrons contribute to the X-ray emission in the
1-5 keV band, and projecting in the plane of the sky.
 This is shown in the bottom-left panel of
Figure 2, which suggests that for cocoons larger than $100$ kpc, typically 
a patch as large as tens of arcseconds would be X-ray bright. 
If one estimates the
fraction of the cocoon area that is X-ray bright, this fraction initially
rises (till $t_j$) to $\sim 0.3-0.5$ of the total area (assuming the
cocoon the projected in the plane of the sky), and then drops.

This calculation of the angular size of the X-ray bright region of
the cocoon allows us to estimate an average
X-ray surface brightness of the cocoon, and we show the results
in the top-right panel of Figure 2, for the same cases as mentioned
earlier. The surface brightness (in the 1-5 keV band) initially drops
in a gradual manner, owing to two competing effects: increasing X-ray luminosity
and an increasing portion of the cocoons which are illuminated by X-ray.
But the it drops rapidly after $t_j$, especially
at high redshift. 

Another important probe of X-ray bright cocoons is the X-ray spectral
index. We calculate the photon index $\Gamma (=\alpha + 1)$ 
in the 1-5 keV band for the same
cases, and the results are shown in the bottom-right panel of Figure 2.
We also include  data points for a number of cases which have been
observed long enough for spectrum determination. 
We have
estimated the cocoon angular size from visual inspection from the observations
cited in the caption, assuming the cocoons to be projected in the plane of the
sky, and estimated the physical size assuming a $\Lambda$-CDM cosmology with
$h=0.7, \, \Omega_0=0.3, \Omega_\Lambda=0.7$.
The error bars
on the data points are large, but there appears to be a trend of 
decreasing photon index with increasing cocoon size: larger cocoons
show flatter X-ray spectrum. 

Our model
predictions show that initially the X-ray spectrum gradually gets flatter 
(smaller value of $\Gamma$) till $t\sim t_j$. This is because of the
fact that for electrons $\gamma \ge 10^3$, the injection of fresh particles
mitigates the radiative steepening of electron spectrum.
The injection of electrons from the jet compensates
for synchrotron energy losses for electrons with $\gamma \sim 10^3$, where
 two relevant time scales become comparable: (a)
the jet lifetime ($10^{7\hbox{--}8}$ yr) 
and the (b) inverse Compton cooling time scale of electrons with 
$\gamma \sim 10^3$ ($\sim 3 \times 10^7$ yr at present cosmological epoch).
This energy scale (corresponding to $\gamma \sim \sqrt{1 \, {\rm keV} \ 
kT_{CMB}} \sim 2 \times 10^3$) also happens to be
the one required to upscatter CMB photons to keV range.
The concurrence of these two timescales 
produces a temporary hardening of the soft X-ray spectrum.

Then, after the jet stops, the number of electrons at this
energy scale rapidly decreases because of inverse
Compton and adiabatic loss, and the spectrum becomes soft again. This
softening occurs rapidly at high redshift, as expected from increasing
inverse Compton loss. With regard to some of the data points with 
large cocoons and $\Gamma \sim 1.5$, we found that they can be explained
with models using a jet power of order $10^{46}$ erg s$^{-1}$ at low redshifts.

\begin{table}
\vskip 0.2in
\begin{center}
\begin{tabular}{|c||c||c||c||c||c|}
\hline
Model & $Q_j$ (erg s$^{-1}$) & $z$ & $t_j$ (yr) & $\Lambda$ (g cm$^{-1})$
 & $\gamma_{min}$  \\
\hline
\hline
Case I & $10^{45}$ & $0.1$ & $10^8$ & $10^{19}$ & $1$\\
\hline
Case II & $10^{46}$ & $0.1$ & $10^8$ & $10^{19}$ & $1$\\
\hline
Case III & $10^{46}$ & $0.2$ & $10^8$ & $10^{19}$ & $1$\\
\hline
\hline
\end{tabular}
\end{center}
\caption{The values of different parameters used in the models cited in the 
text are tabulated here. Apart from these parameters, all models use $\beta=2$, $\Gamma_c
=4/3, R=2$.
}
\end{table}

\subsection{Variations with parameters}
Although we have shown the results for cocoons for a few cases,
varying parameters such as the jet power, redshift and jet life time,
there are other free parameters in this model whose effects must be
understood. The X-ray properties of cocoon also depend on the ambient
density (here parametrized as $\Lambda$) and the lower cutoff in the electron
energy distribution ($\gamma_{min}$). 
To study the effect
of these parameters, we first plot in Figure 3 (with solid lines) the results
for a fiducial case (Case III): 
$Q_j=10^{46}$ erg s$^{-1}$, $z=0.2$, $t_j=10^8$ yr,
and $\Lambda=10^{19}$ g cm$^{-1}$, $\gamma_{min}=1$. 
In this
section we will study the effect of changing the values of $L_j, \Lambda,
\gamma_{min}$ separately, and compare with the results for our fiducial
case. We will not vary the redshift and the jet life time, as the
discussion in the 
previous section has already 
considered the effects of changing these parameters.
Also, we keep $\gamma_{max}$ fixed at $10^6$, and we do not change
the values of $p$ and $r$.

In Figure 3, we have plotted the X-ray flux (in 1-5 keV band; top-left
panel), ratio between X-ray and radio power (top-right panel), X-ray
surface brightness (in 1-5 keV; bottom-left panel), and the photon
index ($\Gamma$) 
in the bottom-right panel. All these parameters are plotted against the
cocoon size. 

First, the results for the fiducial case mentioned above are shown
with solid lines in all panels for easy comparison. Then, with dotted line,
we show the results of changing jet power to $10^{47}$ erg s$^{-1}$.
As expected this cocoon expands to occupy a large volume, and is more
X-ray bright than the fiducial case. 
The ratio of X-ray to radio power does not change, however, but the surface
brightness increases.
The photon index continues
to decrease and the X-ray spectrum becomes harder with time (and size) 
for $t \sim t_j$ after which it becomes softer.

Next, we change the value of $\gamma_{min}$ to $10^3$ 
(keeping all other values same
as in the fiducial case), and the results are shown with short-dashed lines.
Increasing the lower cutoff results in increasing the number density
of electrons for a given total energy in the cocoon, and  produces a
larger X-ray flux and surface brightness. But since there are more electrons 
with high energy in this case compared to the fiducial case, they also
lose energy faster, and consequently, the drop in X-ray power after the
stopping of the jet is faster than in the fiducial case. 
Interestingly, the increasing of $\gamma_{min}$ does not effect the the X-ray
to radio power ratio and the X-ray photon index. (It would, if the 
lower cutoff were to increase beyond $\gamma\sim 10^3$.)

Finally, we change the ambient density to $\Lambda=5\times 10^{19}$ g 
cm$^{-1}$ that is indicative of a dense environment. The results are shown with
long-dashed lines, and they show that the radio cocoon is constrained to remain
small in size in this case. This results in not allowing the cocoon
pressure (and magnetic field) to decrease fast enough, so that electrons
are made to lose energy faster through synchrotron radiation. Both X-ray
and radio power drop rapidly in this case and the ratio between X-ray
to radio power also decreases (although, since the curves are drawn as functions
of $D$, the rapid decrease is not apparent from the figures). 
The photon index also hardens faster than
in the fiducial case. This implies that it would be difficult to detect
X-ray bright cocoons in dense environments like galaxy clusters, although
it would be difficult to observe them anyway because of the bright
X-ray emission from the intracluster gas.

\begin{figure}
\centerline{
\epsfxsize=0.5\textwidth
\epsfbox{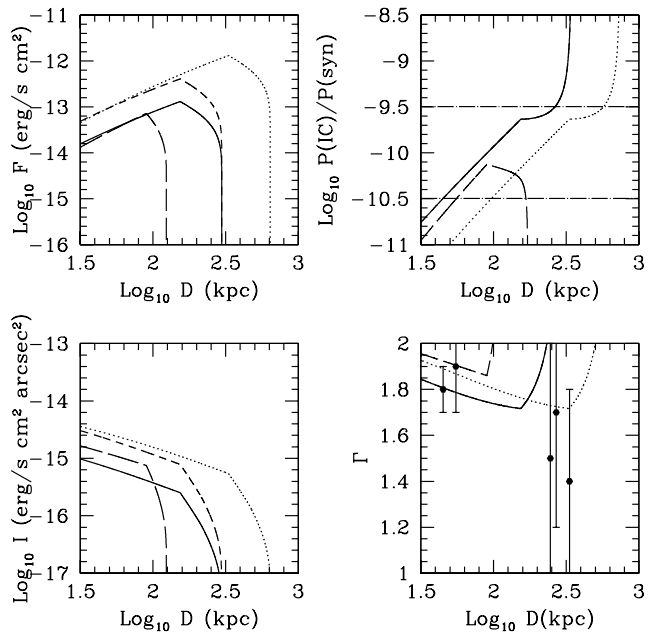}
}
{\vskip-3mm}
\caption{
Evolutions of the X-ray flux (in 1-5 keV band; top-left panel), ratio of
X-ray to radio power (top-right), X-ray surface brightness (bottom-left)
and X-ray photon index (bottom-right) are shown as functions of the 
cocoon size for a few cases. The solid line shows the results for
Case III: $Q_j=10^{46}$ erg s$^{-1}$,
$\Lambda=10^{19}$ g cm$^{-1}$ at $z=0.2$, with $\gamma_{min}=1$, and 
$t_j=10^8$ yr. The dotted line shows the case for $Q_j=10^{47}$  erg s$^{-1}$,
keeping other parameters fixed. The short-dashed lines refer to the
case with $\gamma_{min}=10^3$, and the long-dashed lines show the 
effect of changing the ambient density to $\Lambda=5 \times 
10^{19}$ g cm$^{-1}$. Data points from Figure 2 for 
photon index are again plotted here. 
}
\end{figure}

The X-ray fluxes shown in the top-left panel of Figure 3 suggest that
for typical radio galaxy parameters as used here, the cocoons shine in
soft X-rays with a flux of $1\hbox{--}10$ micro-Crab, or,
$10^{-14}\hbox{--}10^{-13}$ erg cm$^{-2}$ s$^{-1}$ per keV,
before dropping to lower values after
$t_j$.

Curves in the top-right panel of
Figure 3 show that the ratio of X-ray to radio power
is a  robust parameter, in the sense that it does not vary with changes in 
$\gamma_{min}$ and jet power at fixed times. The curves are plotted against$D$, but the kink in the curves due to the stopping of jet occurs for all curves
at the same time, namely, $t_j$, and a comparison of the solid, short dashed
(which happens to be superposed on the solid curve) and the dotted curves
show that the ratio is a robust parameter. 
It does vary with redshift (as we have
seen in Figure 1) and with ambient density (long dashed line in
Figure 3). This robustness allows us
to use this parameter to characterize the connection between radio and X-ray
properties of cocoons, and we will discuss this issue further in the
next section. 

\subsection{Time-averaged ratio of X-ray to radio power} 
Our results show that many of the X-ray properties of FR-II radio
galaxy cocoons vary substantially with time, even when all other
parameters such as jet power, ambient density and others are kept
constant. It is therefore not easy to predict the X-ray properties
of these sources that can be tested with observations, because it is 
difficult to determine the age of these sources from radio or other
observations. 

One can however use the fact that
the evolutionary time-scale of radio galaxies, of order a few hundred
Myr, is much shorter than the Hubble time, even at redshifts when the
radio galaxy population peaked in number density
($z \sim 2$). This implies that one
can use time-averaged quantities related to X-ray emission, and speculate
upon their average properties that could be observed and tested. 

In particular, we would like to determine the average property
of the ratio of luminosities
in X-ray and radio frequencies, $\nu_x P_x/ \nu_r P_r$ of the sources
under consideration, averaged over time until 
their X-ray or radio emission
drops rapidly: $\int {\nu_x P_x \over \nu_r P_r} dt / \int dt$. Our
results (Figure 3, top-right panel) shows that 
this ratio of luminosities is a robust
parameter, and it varies negligibly with the variations in jet power,
lower cut-off in $\gamma$, although it varies strongly with changes in
ambient density. This ratio is however likely to increase with redshift,
and we wish to determine the scaling with redshift.

We therefore compute the X-ray luminosity $\nu_x P_x$ in the 
1-5 keV band, and we choose a radio frequency of $\nu_r= 151$ MHz to compute
$\nu_r P_r$. We compute a time-average of this ratio, summing over 
the duration in which
the sources are both X-ray and radio bright in these frequencies
(in the source frame).

Celotti \& Fabian (2004) have discussed the significance of this
ratio of X-ray and radio luminosities in the case of
radiation from electrons with a single power law energy distribution. 
In this case, this ratio can be written as,
\be
{\nu_x P_x \over \nu_r P_r}={U_{CMB} \over U_B} \Bigl ({\nu_x \nu_B \over
\nu_r \nu_{CMB}} \Bigr ) ^{1-\alpha} \, (1+z)^{3 + \alpha} \,,
\ee
where $\alpha$ is the radio spectral index, $U_{CMB}$ and $U_B$ are the
energy density in CMB photons (at $z=0$) and magnetic field, respectively.
Also, $\nu_B$ and $ \nu_ {CMB}$ are the gyrofrequency and the peak
frequency of CMB photons at present epoch. For a field strength of
$B=0.1\hbox{--}10$ $\mu$G, the ratio between the
luminosities at 1 keV and 151 MHz
(in the observer frame)
can take values $0.08\hbox{--}300$ at $z=0$, for $p=2.6$, and
$0.08\hbox{--}50$ for $p=2$. The ratio increases with redshift, although 
the increase can be mitigated by the possibility that magnetic field
may change with redshift. Celotti \& Fabian chose a conservative
value of unity for this ratio, at all redshifts, and discussed the possible
implications.

\begin{figure}
\centerline{
\epsfxsize=0.5\textwidth
\epsfbox{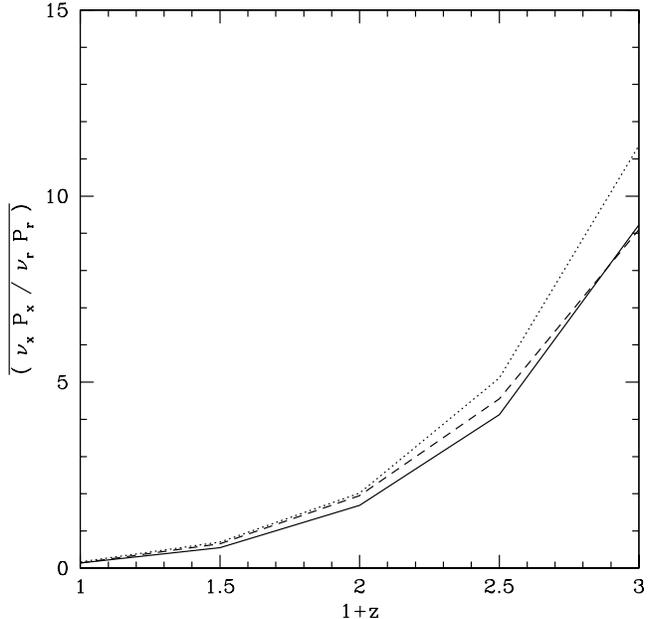}
}
{\vskip-3mm}
\caption{
Time-averaged ratio between X-ray and radio luminosities is shown as a function
of source redshift, for $p=2.5$ (solid line), and $p=2.14$ (dotted line).
The fit ($\sim 0.14 (1+z)^{3.8}$) is shown with a dashed line.
}
\end{figure}

In our model that includes evolving electron populations, 
there is no single power law energy distribution of electrons,
and the magnetic field is estimated through equipartition arguments.
We show the evolution of the 
time-averaged ratio of X-ray and radio luminosities at
the above mentioned frequencies (in the source frame) with redshift in
Figure 4, for two values of $p=2.5$ (solid line), and $p=2.14$ (dotted line),
the last choice of $p$ being motivated by KDA97. Interestingly, the
difference between these two initial values of $p$ is not large, and makes
the time-averaged ratio of luminosities a robust quantity. We find that
the case for $p=2.5$ can be fitted with a simple scaling with redshift,
\be
\overline{(\nu_x P_x/ \nu_r P_r)}_t \sim 0.14 (1+z)^{3.8} \,.
\ee
The results of the two cases of different values of $p$ are different
from the expectation from the simple formula mentioned above. The X-ray
to radio luminosity increases with decreasing $p$ (or increasing $\alpha$), 
especially at high redshift. This is because the increased population of high
energy electrons (for smaller values of $p$) rapidly lose energy, 
decreasing $\gamma$ down to the level where they become X-ray bright, and
this process become more efficient at high redshift.

\subsection{X-ray luminosity function of FR-II radio galaxies}
We are now in a position to estimate the number density of X-ray
bright radio galaxy cocoons using the radio luminosity function of these
sources, and using the above results for the relation between X-ray
and radio emission. We use the radio luminosity function of FR-II 
galaxies, as determined by Willott \etal (2001) from 3CRR, 6CE and
7CRS samples. This luminosity function was determined assuming a 
cosmological model with $\Omega_0=0=\Omega_\Lambda, \, \Omega_k=1, \, h=0.5$.
We have converted it for the $\Lambda$-CDM cosmology with $\Omega_\Lambda=0.7,
\, \Omega_m=0.3, \, h=0.7$ using the relation (Peacock 1985),
\be
\rho_1 (P_1, z) {dV_1 \over dz}=\rho_2 (P_2, z) {dV_2 \over dz} \,
\label{eq:cos}
\ee
where $P$ is the luminosity derived in a specific cosmological model
for a measured flux and at a given redshift $z$, and the indices refer to 
the two different cosmological models. The luminosities in two different
models are related as,
\be
P_1 D_1 ^{-2}= P_2 D_2 ^{-2} \,,
\ee
where 
\be
D (z)={c \over H_0} \int _0 ^z {d z' \over \sqrt{\Omega_M (1+z')^3 +
\Omega _\Lambda}} \,,
\ee
is the comoving distance in a flat cosmological model. Also, the comoving
volume in eqn \ref{eq:cos} is given by $dV=4 \pi D(z) ^2 \, dD(z)$.

Having done this, we have associated a diffuse X-ray luminosity to each FR-II
sources with the scaling result from the last section, and determined
the X-ray luminosity function of FR-II galaxies.

Figure 5 shows the computed X-ray luminosity function, in the units of number
per unit (comoving) $Mpc^3$, per $10^{44}$ erg s$^{-1}$, for three
redshifts: $z=0$ (solid), $z=1$ (dotted) and $z=2$ (short-dashed line).
The peak of the X-ray luminosity function shifts to increasing luminosity
because of the strong redshift evolution of the X-ray to radio power ratio.
We note that
Celotti \& Fabian (2004) assumed a constant ratio of unity and so their
X-ray luminosity function peaked at the same luminosity at different
redshifts (their Figure 3). 

We compare the computed luminosity function 
with that of clusters in X-ray in the
0.5-2 keV band (shown here with long-dashed line)  
as determined from the REFLEX sample by B\"ohringer \etal (2002) (
for a similar cosmology but with $h=0.5$). Recently, Mullis \etal
(2004) found that the cluster X-ray luminosity function decreases with
increasing redshift at the
high luminosity end, but does not evolve significantly below $10^{44}$ erg
s$^{-1}$. The comparison with the luminosity function expected from
FR-II galaxies shows that the number density
of these sources become comparable at the high
luminosities at $z\sim 2$.

We can also estimate the number of diffuse X-ray emitting cocoons in
a given area of the sky above a given X-ray flux limit.
Integrating the above luminosity function to $z\sim 2$, we find that the
numer of 
sources above a flux limit of $\sim 3 \times 10^{-16}$ erg cm$^{-2}$ s$^{-1}$
is of order $\sim 27$ per square degree.
Recently Finoguenov \etal (2010) detected $\sim 6$ X-ray emitting radio lobes
in $1.3$ deg$^2$ above $2 \times 10^{-15}$ erg cm$^{-2}$ s$^{-1}$,
and given the uncertainties, our estimate is consistent with it.
.

\begin{figure}
\centerline{
\epsfxsize=0.5\textwidth
\epsfbox{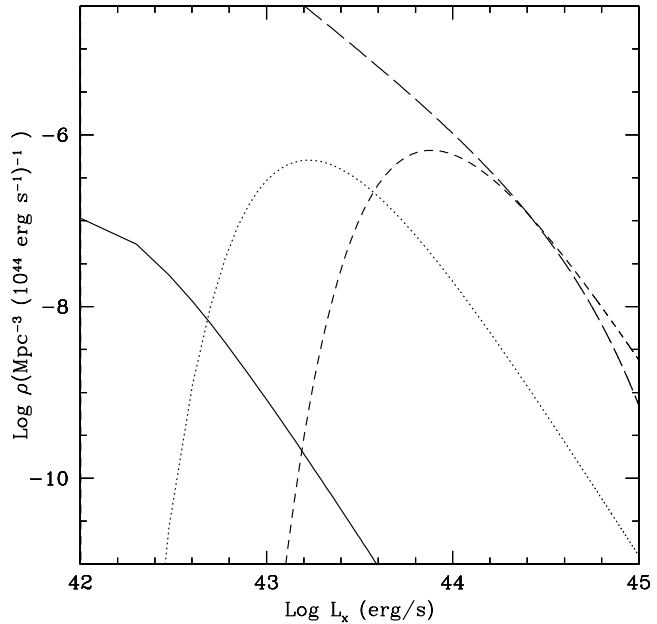}
}
{\vskip-3mm}
\caption{
Predicted X-ray luminosity function of FR-II radio galaxy cocoons,
based on the relation between
X-ray and radio power, is shown here, for $z=0$ (solid), $z=1$ (dotted)
and $z=2$ (short-dashed line). The X-ray luminosity refers to that
in the 1-5 keV band, and the luminosity function shows the comoving
number density of objects, per $10^{44}$ erg s$^{-1}$, in the $\Lambda$-CDM
cosmology, with $h=0.7$. 
The long-dashed line shows the present day X-ray luminosity
function of clusters, based on the REFLEX (Rosat-ESO Flux Limited)
 survey by B\"ohringer \etal (2002).
}
\end{figure}

\section{Discussion}
The expected number of X-ray bright radio galaxy cocoons as estimated
above can be compared with the observed values. Bauer \etal (2002) found
six extended sources in the {\it Chandra} Deep Field North survey,
within an area of $\sim 130$ arcmin$^2$, implying a surface density
of $\sim 167^{+97} _{-67}$ deg$^{-2}$, at a limiting soft X-ray flux
of $\sim 3 \times 10^{-16}$  erg cm$^{-2}$ s$^{-1}$. Our estimate shows
that a small fraction of order $\sim 0.1\hbox{--}0.3$ of such extended
soft-Xray  sources in the sky could be due to FR-II radio galaxies.

The X-ray luminosity function in the soft band can also be used to 
estimate mirror effect of X-ray emission, namely, the Sunyaev-Zeldovich
(SZ) effect on the CMB. To some extent, it 
would underestimate the effect because
of the fact the soft-band X-ray power is smaller than the total X-ray 
power. Keeping this in mind, we can determine the ratio
of the total radiation energy density in
the soft-Xray band that is emitted by these sources and that is present
in the CMB, by integrating
the luminosity function: ${\Delta \epsilon \over \epsilon}
\sim \int {dt \over dz} {dz
\over a T_{CMB,z}^4}
\int L_x {d \rho (L_x, z) \over dL_x} dL_x$. We find that $\Delta \epsilon/
\epsilon \sim 2 \times 10^{-7}$, integrating up to a redshift $z \sim 2$, 
although strictly speaking this is a lower limit.
This implies a Compton y-parameter of
order $y \sim (1/4) {\Delta \epsilon \over \epsilon} \ge  10^{-7}$. 
Yamada, Sugiyama \& Silk (1999) considered the distortion of the CMB
from the population of radio galaxy cocoons, using Press-Schechter mass
function and assuming that halos above a certain mass limit produce radio
galaxies, and estimated that $y\sim 6 \times
10^{-5}$. En{\ss}lin \& Kaiser (2000) however estimated from radio galaxy
luminosity functions a total optical depth
of the relativistic electrons to be $\tau \sim 10^{-7}$. 
For relativistic electrons with $\gamma_{min}
\ge 1$, one has $y_{nth} \ge \tau$, and our estimate is 
consistent with that of En{\ss}lin \& Kaiser (2000).

It is not only the cocoons of radio galaxies that have been 
observed in X-rays, but the jets have also been recently
 detected by {\it Chandra}
(e.g., Schwartz 2002). Although some uncertainties remain in identifying
the emission mechanism behind the X-ray radiation, inverse Compton scattering
of CMB photons is thought to be one of the possibilities (Stawarz \etal 2004).
it is possible that the X-ray radiation from the jets may contaminate 
the emission from the cocoons and make the study of cocoons in X-rays difficult,
especially at high redshift where angular resolution may become an issue.

\section{Conclusion}
We have used an evolving model of FR-II radio galaxy cocoons that takes
into account the stopping of the jet after a jet lifetime of $t_j$,
and we have studied 
the X-ray property of cocoons from inverse Compton scattering of
CMB photons. We have shown that the the X-ray power (in the 1-5 keV band)
and the X-ray surface brightness
decrease rapidly after the stopping
of the jet, as well as the radio power. 
At the same time, 
the X-ray photon index is found first decrease to some extent while the
jet is active. This occurs because at the important energy scale for
inverse Compton scattering of CMB photons to keV range ($\gamma \sim 10^3$),
two relevant timescales become comparable: the jet lifetime and that of
cooling due to inverse Compton ($10^{7\hbox{--}8}$ yr). After the stopping
of the jet, the photon index increases, making the spectrum
softer. We find that typical radio galaxy parameters
produce an X-ray flux (1-5 keV band) of order $1\hbox{--}10$ $\mu$-Crab,
being emitted from regions with sizes of order of tens of arcsec.

We have also shown that
the ratio of X-ray to radio power changes little with jet power and 
lower cut-off in the electron energy distribution, and it 
changes mostly with redshift and ambient density. We have then determined the
time-averaged ratio of X-ray to radio power and studied its scaling with
redshift, which we found to scale as $\propto (1+z)^{3.8}$, almost independent
of the injected electron energy spectrum, contrary to the expectations from
a population of electrons with a single power law distribution of energy.

Using these scaling relations, we estimated 
the X-ray luminosity function of FR-II
radio galaxies at various redshifts, and estimated the number of 
such diffuse X-ray sources to be $\sim 25$ deg$^{-2}$ above an X-ray
flux limit of $\sim 3 \times 10^{-16}$ erg cm$^{-2}$ s$^{-1}$
(compared to $\sim 167$ deg$^{-2}$ diffuse
soft X-ray sources observed in the
{\it Chandra} Deep Field North for a similar flux limit).

\bigskip
I am grateful to Paddy Leahy, the referee, for valuable comments.
I also thank Mitchell Begelman for his comments on the paper and
Biswajit Paul for discussions.

\bsp

\end{document}